\documentclass[manuscript]{aastex}
\usepackage{natbib}
\usepackage{epsfig}
\usepackage{epstopdf}
\usepackage{morefloats}
\usepackage{amsmath}

\newcommand{\ik}{{\it Kepler~}}
\newcommand{\ikt}{{\it Kepler}}
\newcommand{\kepler}{{\it Kepler}}

\shorttitle{Spot-induced TTV - Theory}
\shortauthors{Mazeh, Holczer and Shporer}

\begin{document}

\title{Time variation of \ik transits induced by stellar rotating
spots
--- a way to distinguish between prograde and
retrograde motion I. Theory}

\author{
Tsevi Mazeh\altaffilmark{1,2},
Tomer Holczer\altaffilmark{1},
Avi Shporer\altaffilmark{3, 4, 5}
}

\altaffiltext{1}{School of Physics and Astronomy, Raymond and
Beverly Sackler Faculty of Exact Sciences, Tel Aviv University,
Tel Aviv 69978, Israel}

\altaffiltext{2}{The Jesus Serra Foundation Guest Program,
Instituto de Astrofísica de Canarias, C. via Lactea S/N, 38205 L
 Laguna, Tenerife, Spain}

\altaffiltext{3}{Division of Geological and Planetary Sciences,
California Institute of Technology, Pasadena, CA 91125, USA}

\altaffiltext{4}{Jet Propulsion Laboratory, California Institute
of Technology, 4800 Oak Grove Drive, Pasadena, CA 91109, USA}

\altaffiltext{5}{Sagan Fellow}

\begin{abstract}
Some transiting planets discovered by the {\it Kepler} mission
display transit timing variations (TTVs) induced by stellar spots
that
rotate on the visible hemisphere of their parent stars. An
induced TTV can be observed when a
planet crosses a spot and modifies the shape of the transit light
curve, even if the time resolution of the data does not allow to
detect the crossing event itself.
We present an approach that can, in some cases, use the derived
TTVs of a planet  to distinguish between a prograde and
a retrograde planetary motion with respect to the stellar
rotation.\\
Assuming a single spot darker than the stellar disc, spot
crossing by the planet can induce measured positive
(negative) TTV, if the crossing occurs in the first
(second) half of the transit. On the other hand,
the motion of the spot towards (away from) the center of the
stellar visible disc causes the stellar brightness to decrease
(increase).
Therefore, for a planet with prograde motion, the induced
TTV is positive when the local slope of the
stellar flux at the time of transit is negative, and vice versa.
Thus, we can expect to observe a negative (positive) correlation
between the TTVs and the photometric slopes for prograde
(retrograde) motion.
Using a simplistic analytical approximation, and also the
publicly available SOAP-T tool to produce light curves of
transits with spot-crossing events, we show for some cases how
the induced TTVs
depend on the local stellar photometric slopes at the transit
timings.
Detecting this correlation in \ik transiting systems with high
enough signal-to-noise ratio
can allow us to distinguish between prograde and retrograde
planetary motions.
In coming papers we present
analyses of the KOIs and \kepler\ eclipsing binaries, following
the formalism developed here.
\end{abstract}


\section{Introduction}
\label{intro}

Formation and evolutionary processes of stellar and planetary
systems are
expected to leave their imprint on the present-day systems. One
such imprint is the stellar obliquity, the angle between the
stellar spin axis and the orbital angular momentum axis, also
referred to as the spin-orbit
angle. For star-planet systems the measurement of this angle is
a matter of intense study in recent years
\cite[e.g.,][]{triaud10, moutou11, winn11, albrecht12},
primarily for hot Jupiters ---
gas-giant planets at short-period orbits. Some of the systems
were found to be aligned, in a prograde orbit with spin-orbit
angle close to
zero, while others were found to be
misaligned, including systems in retrograde motion where the
spin-orbit angle is close to $180^\circ$
\citep[e.g.,][]{hebrard11, winn11}.

The growing sample and the wide range of spin-orbit angles
measured for hot Jupiters can be used for studying their
orbital evolutionary history. For example, \cite{winn10} have
noticed that hot stars, with an effective temperature above
6,250 K, tend to have a wide obliquity range, while cool stars
tend to have low obliquities, mostly consistent with well aligned
orbits. This was confirmed by a study of a larger sample by
\cite{albrecht12} and is consistent with the results of
\cite{schlaufman10} and \cite{hansen12} who used different
approaches. Those authors suggested that some mechanisms can
cause
the planetary orbit to attain large obliquity
\citep[e.g.,][]{fabrycky07, naoz11, batygin12}.
Then, tidal  interaction with the host star
\citep[e.g.,][]{winn10} or magnetic braking
\citep[e.g.,][]{dawson14} act to realign the orbit. Since these
processes are probably inefficient for hot stars, those
systems might still retain their wide obliquity range.

So far spin-orbit alignment has been studied primarily through
the Rossiter-McLaughlin (RM) effect \citep{holt1893,
schlesinger1910, rossiter24, mclaughlin24}, originally suggested
for stellar eclipsing binaries, and observed by monitoring the
anomalous radial-velocity signal during eclipse, as the eclipsing
star moves across the disc of the eclipsed star. The RM effect is
sensitive to the sky-projected component of the spin-orbit angle,
and was successfully measured for many transiting planet systems
\cite[e.g.,][]{queloz00, winn06, triaud10}, transiting brown
dwarfs and low-mass star systems \citep{triaud13}, and stellar
binaries \citep{albrecht07, albrecht09, albrecht11, albrecht14}.

The line-of-sight component of the spin-orbit angle can be
measured using
asteroseismology \citep{gizon03, chaplin13}, or the observed
 rotational broadening of spectral lines, {\it if} the
host star radius and rotation period are known with sufficient
precision \citep[][see also
\citealt{schlaufman10}]{hirano12, hirano14}.
However, these two methods require obtaining new data for each
target, using valuable resources (e.g., large telescopes or \ikt\
short-cadence data).
Other methods have been presented, based on stellar gravitational
darkening \citep{barnes09, szabo11, barnes11}, and the beaming
effect \citep[Photometric RM ---][]{shporer12, groot12}.

An interesting
approach was taken by \cite{nutzman11} and \cite{sanchis11a}, who
use the brief photometric signals during transit induced by the
transiting object moving across spots located on the surface of
the host object.
%
This is based on the fact that many stars show photometric
modulations resulting from the
combination of stellar rotation and non-uniform longitudinal
spots distribution \citep[e.g.,][]{irwin09, hartman11,
mcquillan14}.
When such a star displays transits by an orbiting planet, the
transiting object might
momentarily eclipse a stellar spot, inducing an increase in
observed flux, if the surface brightness of the spot-covered area
is lower than that of the non-spotted areas.
%
The derivation of the stellar obliquity requires identification
of such `spot-crossing' events
within a few transits, and estimate the spot and the planet
phases within their motion over the stellar disc.
The method has since
been applied to additional systems using {\it high-speed} \ik and
CoRoT data
\citep{sanchis11b, desert11, deming11, sanchis12, sanchis13}.

We present here another version of this approach that does not
require such high-speed photometry. Instead, we use the fact that
a spot-crossing event can induce measurable transit time
variation
\citep[TTV; e.g.,][]{sanchis11a, fabrycky12, mazeh13, szabo13,
oshagh13b},
even for data that cannot resolve the event itself. Our approach
relies on the expected correlation between the induced TTV and
the
corresponding local photometric slope immediately outside the
transit, presumably induced by the
same spot. Detected correlation or anti-correlation between the
TTVs and
their local slope can in principle differentiate between prograde
and retrograde rotation of the primary star in
stellar binaries or star-planet systems.

 We present here the
basic concept and develop an analytical simplistic approximation
for the
induced TTVs and the photometric slope.
We also use the work of \citet{boisse12} and \citet{oshagh13a},
who developed a numerical tool ---
SOAP-T\footnote{http://www.astro.up.pt/resources/soap-t/}, to
simulate
a planetary transit light curve which includes a spot-crossing
event.
\citet{oshagh13b} used SOAP-T to derive detailed transit light
curves, and then fitted them with transit templates to obtain the
expected TTVs, very similar to what is performed when deriving
the TTVs from the \kepler\ actual data \citep[e.g.,][]{mazeh13}.
We show that our approximation yields TTVs with the same order of
magnitude as the results of \citet{oshagh13b}.
Using our approximation and the SOAP-T tool
we show that in some cases we can expect a negative (positive)
correlation between the
 TTVs induced by spot crossing and the local photometric slopes
at the transit
timings for  prograde (retrograde) motion of the planet.
We also discuss the limitations of this approach when applied to
real data, showing that it can be applied only to a limited
number of systems.

The paper is organized as follows. Section~\ref{sec:principles}
outlines the basics of our approach, while Section~\ref{sec:ttv}
presents the analytical approximation for the induced TTV for
different cases, and Section~\ref{sec:simulations} compares our
approximation with numerical simulations we performed and those
of \citet{oshagh13b}.
In Section~\ref{sec:slopes} we derive the expected derivative of
the stellar brightness at the time of transit, and in
Section~\ref{sec:correlations} display the expected correlation
between the induced TTVs and the stellar photometric slopes.
Finally,
Section~\ref{sec:discussion} discusses our results and the severe
limitations of its applicability to real data.

The present paper is the first of three studies. The next study
(Holczer et al., in preparation)
presents our analysis for the \ik planet
candidates \citep{batalha13}. In that paper we show that indeed a
few systems do show highly significant correlation between their
derived TTVs and the local photometric derivatives, as predicted
by  this work. A forthcoming paper will
present our analysis of the \ik eclipsing stellar binaries
\citep{slawson11}.

\section{The principle of the approach}
\label{sec:principles}

To present our approach, we consider a transiting planet that
crosses
 a stellar spot during its apparent motion over the stellar disc.
Let us assume, for the sake of simplicity, that only one spot is
present on the
stellar disc and that the stellar rotation and orbital axes are
parallel to each other. This includes both prograde (complete
alignment, with obliquity of $0^{\circ}$)
and retrograde (obliquity of $180^\circ$) configurations.
The {\it sign} of the induced time
shift depends on whether the spot-crossing event occurs in the
first (positive TTV) or second (negative TTV) half of the
transit, which is determined by
the location of the spot on the stellar disc at the time of
transit.
%
rotation

\begin{figure}
\includegraphics[width = 0.9\textwidth]{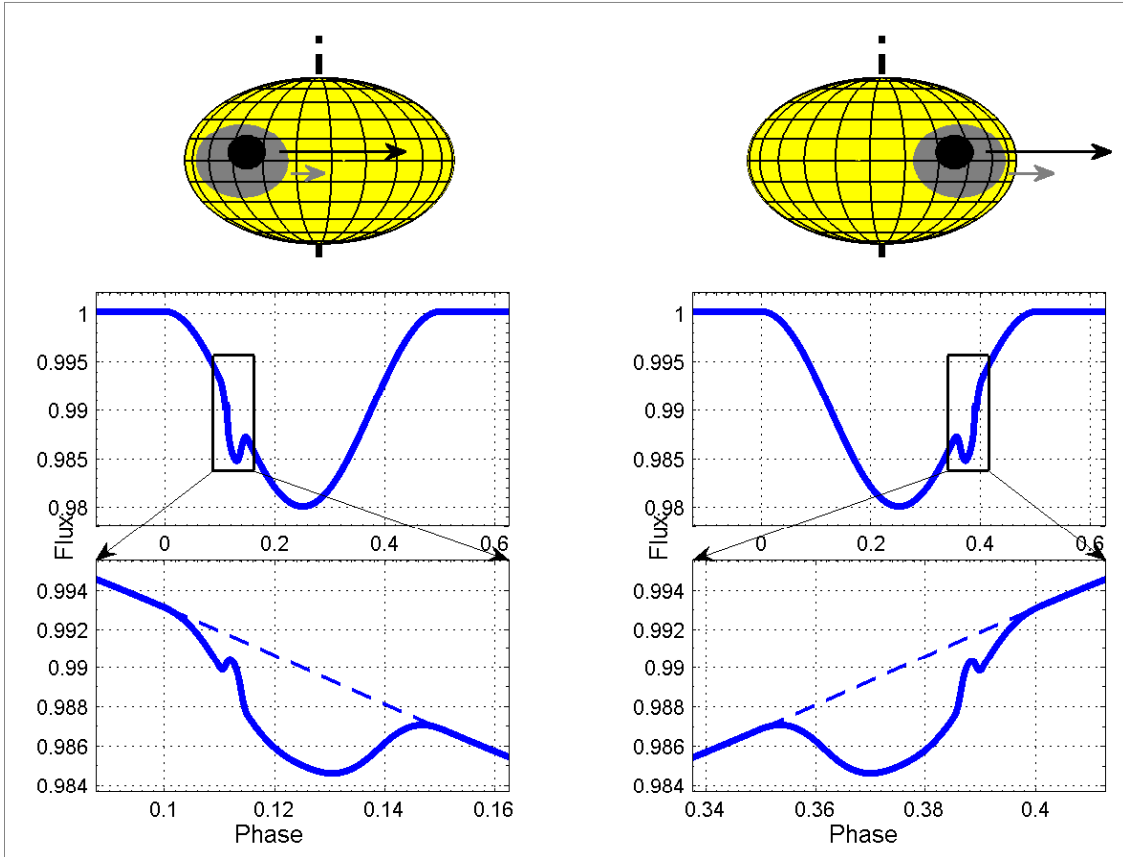}
\caption{Prograde motion --- spot-crossing events during the
first (left) and second (right) halves of the transit. The top
panels display the stellar visible disc (yellow), the planet
(black, small) and the spot (gray, large). The arrows represent
the direction and speed of
the planet and spot relative to the observed stellar disc. The
middle
panels show the light curve due to the spot passage over the
stellar disc, spanning half a stellar rotation period. In the
middle panels we also see the transits, occurring at
phase 0.13 (left) and 0.37 (right) of the stellar rotation. The
bottom panels show the light curves again, now zooming on the
transits, where the small `bumps' are caused by the
spot-crossing events. We consider only a single spot, so the flux
is equal to unity when the spot is on the stellar hemisphere
hidden from the observer's view.
{\it Left (right)} --- the spot is at the first (second) half of
its crossing over the stellar disc, and therefore the local
photometric slope is negative (positive). The planet is at the
first (second) half of the transit and therefore the derived
transit timing shift (while the spot-crossing is unresolved) is
positive (negative).
}
\label{fig:prograde}
\end{figure}

\begin{figure}
  \includegraphics[width = 0.9\textwidth]{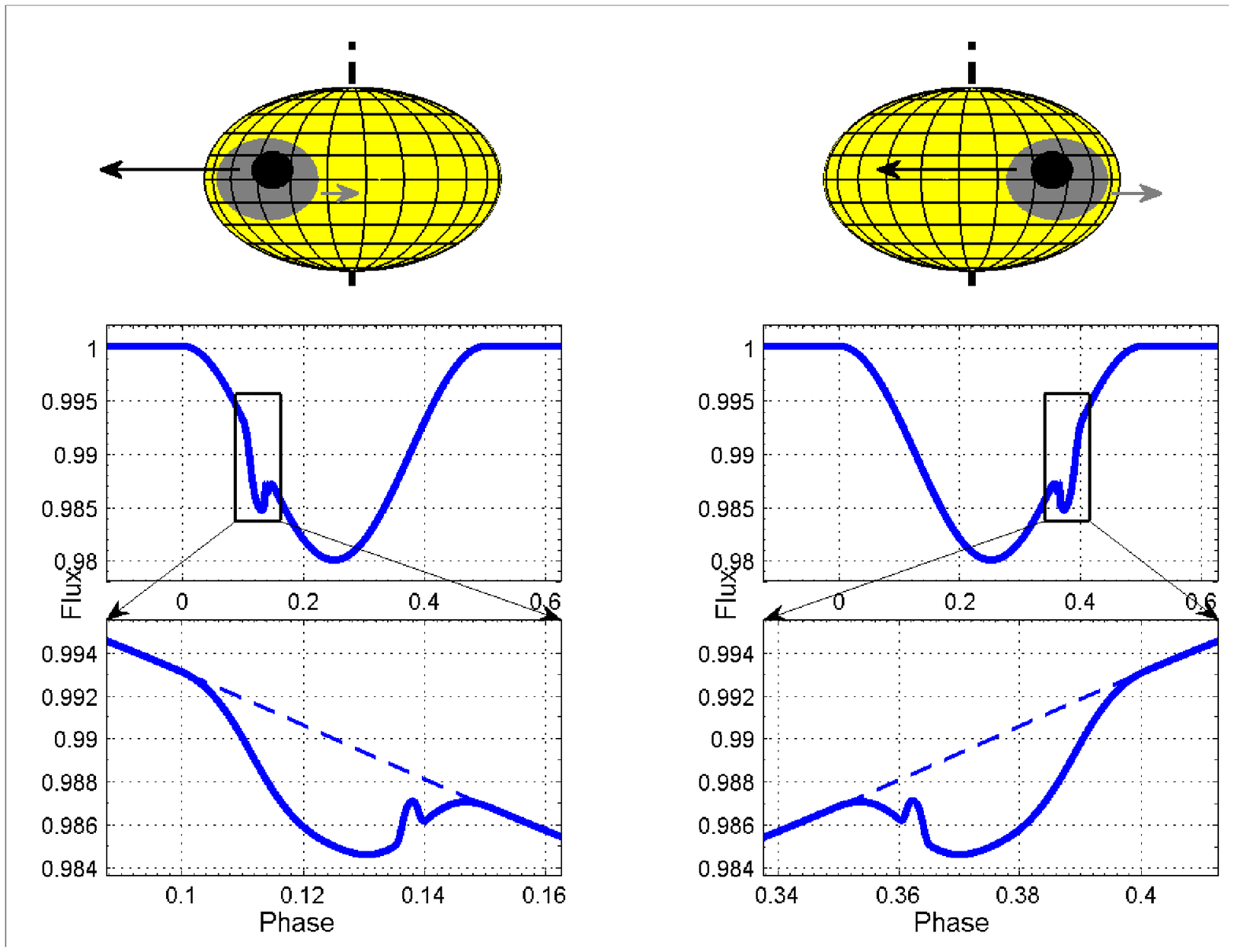}
   \caption{Retrograde motion --- see Figure~\ref{fig:prograde}
for details.
{\it Left (right)} --- the spot is at the first (second) half of
its crossing over the stellar disc, and therefore the local
photometric slope is negative (positive). The planet is at the
second (first) half of the transit and therefore the derived
transit timing shift (while the spot-crossing is unresolved) is
negative (positive). }
\label{fig:retrograde}
\end{figure}

As depicted in Figures~\ref{fig:prograde} \&
\ref{fig:retrograde},
the location of the spot over the stellar
disc determines whether the star is becoming brighter or dimmer
at the time of transit. When the spot is
moving towards (away from) the center of the disc
the stellar intensity is decreasing (increasing), because of the
aspect effect, which changes the effective area of the spot on
the stellar visible disc ---
the projected area of the spot onto the sky plane.
Therefore, when the spot is on the disc edge
its effective area is minimal. On the other hand, the effective
area reaches its maximum when the spot is at its closest position
to the center of the visible disc, when the stellar surface is
(almost) perpendicular to our line of sight.
Another phenomenon, which also causes the star to become fainter
when the spot moves towards the center of the disc is the
limb-darkening effect,
which is ignored at this point of the discussion.

Now, when the stellar rotation and planetary motion have the same
sense of rotation, the spot-crossing event in the first (second)
half of the transit should always occur when the spot is moving
towards (away from) the center of the disc.
Therefore the {\it signs} of the induced TTV and the {\it slope}
of the stellar brightness at the time of transit should be
opposite. This is depicted in Figure~\ref{fig:prograde} for
prograde motion. For retrograde
motion, positive TTV should be associated with positive
slope, as depicted in Figure~\ref{fig:retrograde}.

Therefore, we expect negative correlation between the derived
TTVs and the corresponding stellar photometric slopes for a
system with planetary prograde motion and positive correlation
for a system with retrograde motion.
In the next sections we will show that this is indeed the case
for a limited number of cases by deriving analytical
approximations for the TTVs and the photometric derivatives and
by numerical simulations for the TTVs.

\section{Analytical approximation for the TTV induced by the
spot-crossing event}
\label{sec:ttv}
\subsection{Center-of-light approximation}

\begin{figure}
\centering
\resizebox{15cm}{15cm}
{\includegraphics{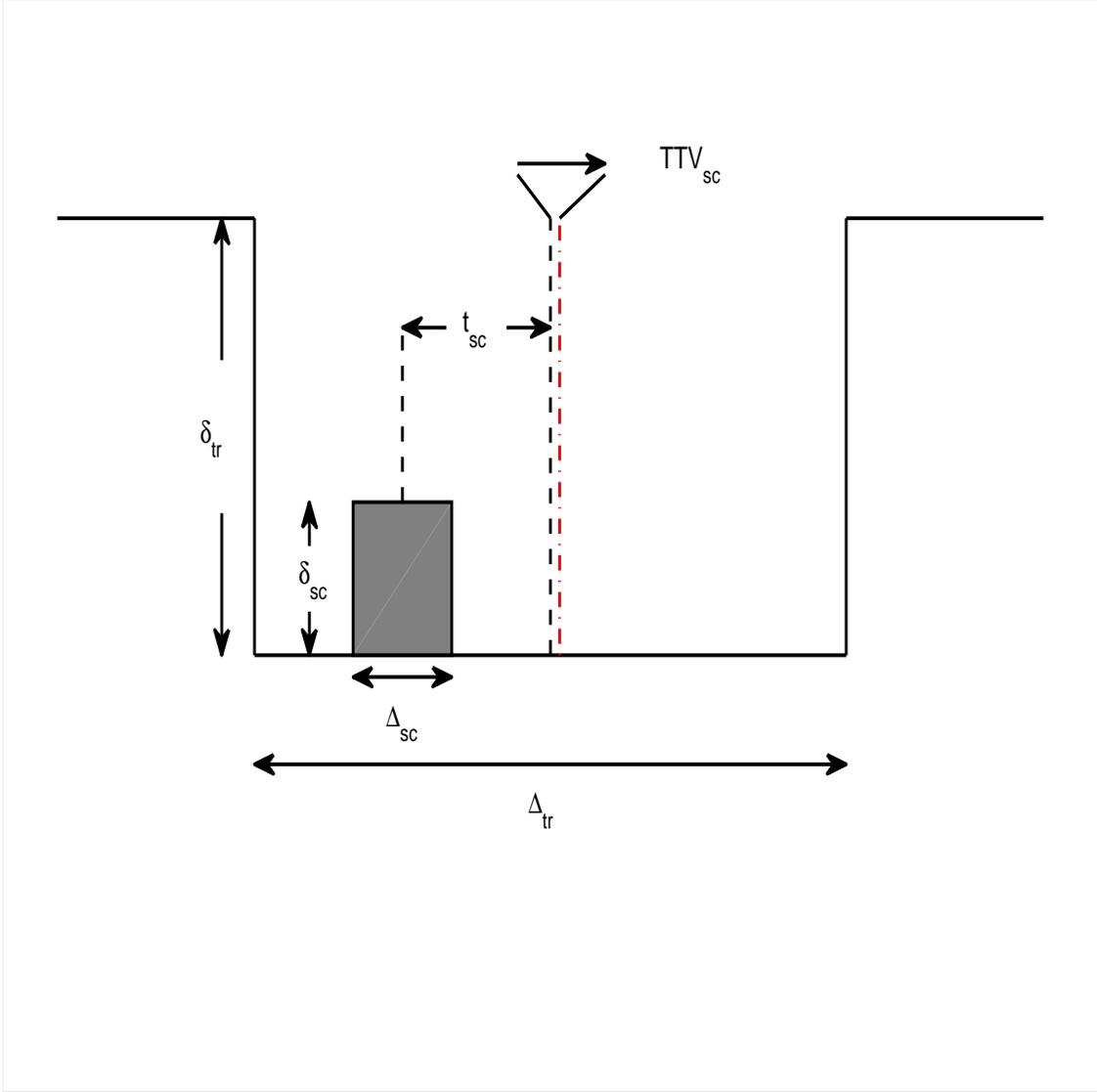}}
\caption{Schematic diagram of a transit light curve with
a spot-crossing event. The transit depth is $\delta_{tr}$, while
$\delta_{sc}$ is the flux increase due to the spot-crossing
event, $\Delta_{tr}$  and $\Delta_{sc}$ are the transit and
spot-crossing durations, and $t_{sc}$ is the timing of  the
spot-crossing event relative to
the mid transit.
The vertical dashed black line represents the expected
mid-transit timing, without any spot-crossing event, while
the red dash-dot line represents the new mid-transit measurement,
due to the shift induced by the spot crossing.
The difference between the two lines, $TTV_{sc}$, is the induced
TTV. Approximately, $TTV_{sc}\simeq - t_{sc} \times
\delta_{sc}\Delta_{sc}/
(\delta_{tr} \Delta_{tr} - \delta_{sc} \Delta_{sc}$).
}
\label{fig:drawing}
\end{figure}

To present the concept behind our method in a more quantitative
way, Figure~\ref{fig:drawing}
shows a simplified
schematic diagram of a transit light curve with a single
spot-crossing event. We neglect the transit
ingress and egress finite duration of both the transit and the
spot-crossing event.
We also assume that at the time of
each transit there is only one circular spot on the stellar disc.
In the figure we also
neglect the limb-darkening photometric modulation, and will
consider this effect later.

In the figure, $\delta_{tr}$ and $\delta_{sc}$ are the
depth of the transit and the amplitude of the photometric
increase inside the transit due to the spot-crossing event,
respectively,
$\Delta_{tr}$ and $\Delta_{sc}$ are the duration of the
transit and the spot-crossing event, respectively, and
$t_{sc}$ is the timing of  the spot-crossing event relative to
mid-transit time.

From the figure one can see, using `center-of-light'
formulation, that we expect the TTV induced by the spot-crossing
event to be:
\begin{equation}
TTV_{sc} \simeq - t_{sc} \frac{\delta_{sc} \Delta_{sc}}
{\delta_{tr} \Delta_{tr} - \delta_{sc} \Delta_{sc}} \ .
\label{eq_timing}
\end{equation}
%
%
This result is similar to Equation (3) of \citet{sanchis11a}
after neglecting $\delta _{sc} \Delta_{sc}$ in the denominator.
We will adopt this approximation below.

\newpage

\subsection{Basic Model}

To derive the analytical simplistic model we first consider a
case for which
\begin{itemize}
\item
the impact parameter of
the spot and the planet are both equal to zero, namely that they
both cross the center of the stellar disc, and
\item
there is no limb darkening.
\end{itemize}
We lessen these two assumptions below.

We denote the location of the spot on the stellar disc by the
angle $\psi$, which is the angle
between the observer and the spot, as seen from the stellar
center. {\it If} the motion of the spot is equatorial, then
$\psi$ is the longitude of the spot on the stellar visible
hemisphere:
\begin{equation}
 \psi(t)=\omega_{*}t \, ,
\end{equation}
where $\omega_{*}$ is the stellar angular
velocity. When the spot is on the stellar limb
entering the visible hemisphere, $\psi$ gets the value of
$-\pi/2$, and when the location of the spot is in the middle of
its visible chord $\psi=0$.

We denote the angle corresponding to the spot crossing by
$\psi_{sc}$. The sky-projected distance of the spot from the
stellar center, as seen by the
observer, is
$d_{sc}= R_*\sin \psi_{sc}$,
where $R_*$ is the stellar radius. The timing of the
spot-crossing event,
measured relative to the middle of the transit, is therefore
%
\begin{equation}
 t_{sc}=  \frac{\Delta_{tr}}{2}\sin \psi_{sc}   \, .
\label{eq_t_sc}
\end{equation}
%
%

In order to estimate the induced TTV,
let us consider two extreme cases: a small spot, for which
\begin{equation}
R_{spot}\ll R_{pl}\ll R_* \, ,
\end{equation}
and a large spot, for which
\begin{equation}
R_{pl}\ll R_{spot}\ll R_* \ ,
\end{equation}
where $R_{pl}$ and $R_{spot}$ are the radii of the planet and the
spot, respectively.

For both cases we introduce a darkness parameter, $0<\alpha<1$,
which
measures the surface brightness of the spot relative to the
surface brightness of the star immediately outside the spot. A
completely dark spot would have $\alpha=0$, while
$\alpha$ close to unity means the spot is almost as bright as the
unspotted stellar area.

For the small-spot approximation we can assume that the spot is
completely covered  by the planet during the spot-crossing event
and therefore
\begin{equation}
{\rm Small \  spot\!: \ \ \ \
\ }\delta_{sc}\simeq(1-\alpha)\left(\frac{R_{spot}}{R_*}\right)^2
\cos\psi_{sc}
\ \ , \ \
 \Delta_{sc}\simeq\Delta_{tr}\left(\frac{R_{pl}}{R_*}\right)\, .
\label{small_basic}
\end{equation}
As noted above, the factor $\cos\psi_{sc}$ comes from the fact
that the effective area  of the spot  is reduced by the aspect
ratio, which is a function
of the spot position on the visible stellar disc.

%
%
For the large-spot approximation the planet is fully contained in
the spotted area during the spot-crossing event, and therefore we
get
\begin{equation}
{\rm Large \ spot\!: \ \ \ \ \ }
\delta_{sc}\simeq(1-\alpha)\left(\frac{R_{pl}}{R_*}\right)^2
\ \ , \ \
 \Delta_{sc}\simeq\Delta_{tr}\left(\frac{R_{spot}}{R_*}\right)
\cos\psi_{sc} \, .
\label{large_basic}
\end{equation}
Here the factor $\cos\psi_{sc}$ comes from the fact that the time
to
cross the spot by the relatively small planet is reduced by the
same
aspect ratio.

We now approximate the TTV to be
\begin{equation}
TTV_{sc} \simeq - t_{sc}
\frac{\delta_{sc}}{\delta_{tr}}\,
\frac{\Delta_{sc}}{\Delta_{tr}}\, ,
\end{equation}
and the transit depth $\delta_{tr}$ to be on the order of
$(R_{pl}/R_*)^2$.  We therefore get for the small-spot
approximation
\begin{equation}
\begin{split}
{\rm Small \  spot\!: \ \ \ \ \ }
TTV_{sc}
&
\simeq  - t_{sc} (1-\alpha)\left(\frac{R_{spot}}{R_*}\right)^2
\frac{R_{pl}}{R_*} \left(\frac{R_{pl}}{R_*}\right)^{-2}
\cos\psi_{sc}
\\&
=- t_{sc}(1-\alpha)\frac{R_{spot}^2}{R_{pl}R_*}\cos\psi_{sc}
\, ,
\end{split}
\label{eq:timing_small1}
\end{equation}
%
%
and for the large-spot approximation
%
\begin{equation}
{\rm Large \  spot\!: \ \ \ \ \ }
TTV_{sc}
\simeq  - t_{sc} (1-\alpha)   \frac{R_{spot}}{R_{*}}
\cos\psi_{sc}
\ .
\label{eq_timing_large1}
\end{equation}
Note that when $R_{spot}\rightarrow R_{pl}$,
Equation~(\ref{eq:timing_small1}) $\rightarrow$
Equation~(\ref{eq_timing_large1}). To ease the discussion we
define
$\mathcal{R}$ as:
\begin{equation}
\mathcal{R} = \left\{
\begin{array}{l l}
\frac{R_{spot}^2}{R_{pl}R_*} & $for small spot$\\
\frac{R_{spot}}{R_*}               & $for large spot$ \,.
\end{array} \right.
\label{eq_factor2}
\end{equation}
Using Equation~(\ref{eq_t_sc}) we get:
%
%
\begin{equation}
TTV_{sc}
\simeq
-(1-\alpha) \mathcal{R}
\frac{\Delta_{tr}}{2}
\cos\psi_{sc}\sin\psi_{sc}
\ ,
\label{eq_TTV}
\end{equation}
which is valid both for the small- and large-spot approximations.
The maximum observed TTV induced by the spot crossing is
\begin{equation}
{\rm max}\{TTV_{sc}\}
\simeq
\frac{(1-\alpha)}{4}\mathcal{R}\,
\Delta_{tr}
\ .
\label{eq_timing_max}
\end{equation}

\subsection{Models for Limb darkening, impact parameter and
obliquity}
\label{limb_darkening}

\subsubsection{Limb darkening}
%
To include the stellar limb darkening effect in our model, we
consider a  quadratic limb-darkening law of
$\mathcal{S}=1 - g_1(1-\cos \psi) - g_2(1-\cos \psi)^2 $, where
 $\mathcal{S}$ is the scaled stellar surface brightness and $g_1$
and $g_2$
are the two limb-darkening coefficients, such that $g_1 + g_2
<1$.

The induced TTV is proportional
to $\delta_{sc}$, the increase of the stellar brightness during
the spot crossing, which depends linearly on
the  stellar surface brightness $\mathcal{S}$, which is now a
function of $\psi$. Therefore we get
\begin{eqnarray}
TTV_{sc} =-(1-\alpha) \mathcal{R}\frac{\Delta_{tr}}{2}
\cos \psi(t) \sin \psi(t) \,
\big\{1 - g_1(1-\cos \psi) - g_2(1-\cos \psi)^2\big\} \nonumber
\\
=-(1-\alpha) \mathcal{R}\frac{\Delta_{tr}}{2}
\big\{(1-g_1-g_2) \sin \psi(t) + (g_1 + 2g_2) \sin \psi (t) \cos
\psi (t) \nonumber \\
 - g_2 \sin \psi (t) \cos^2 \psi (t) \big\}
\cos \psi(t) \, .
\label{eq:TTV_basic_eq_LD}
\end{eqnarray}

Note that because of the limb darkening the transit light curve
does not have a rectangle shape, so our Equation (1) should be
modified. Nevertheless, as this analytical approach is aimed only
to understand the features of the TTVs as a function of the
spot-crossing phase, we neglect this effect that will affect all
phases alike.

\subsubsection{Impact parameter}

Another extension of our simplistic model accounts for a
non-zero impact parameter, $b=\cos\theta$.
Note that the stellar rotation is, as before, orthogonal to our
line of sight.
In this extension of the simplistic model, both
planet and spot still have the same impact parameter, namely both
move along the same chord on the stellar disc, a chord that does
not go through the center of the disc.
Therefore, the spot moves at a colatitude $\theta_{spot}=\theta$,
with an impact parameter  $b_{spot}=\cos\theta_{spot}$.
In such a case, the angle $\psi$ fulfill the relation
\begin{equation}
\cos\psi=\sin\theta\cos\phi \, ,
\end{equation}
where now $\phi$ is the longitude of the planet, and  $\phi=0$ is
when the planet crosses the projection of the stellar rotational
axis.
The range of $\psi$ is now
different: $ \pi/2-\theta\leq|\psi|\leq\pi/2$, and the timing of
the spot crossing is
\begin{equation}
t_{sc}=  \frac{\Delta_{tr}^b}{2}\sin \phi_{sc}\, ,
\end{equation}
where $\Delta_{tr}^b$ is the transit duration when $b\neq0$. A
good approximation would be
$\Delta_{tr}^b= \Delta_{tr}\sin \theta$.

We now separate the discussion for the small and large spot
approximations.
For small spot, the duration of the spot-crossing event,
$\Delta_{sc}$, is still
the same as for $b=0$, but the transit duration is shorter by a
factor of $\sin\theta$. The flux increase depends on $\cos\psi$,
as for $b=0$.  We can therefore write
%
\begin{equation}
{\rm Small \  spot\!: \ \ \ }
\delta_{sc}\simeq(1-\alpha)\left(\frac{R_{spot}}{R_*}\right)^2
\sin\theta\cos\phi_{sc}
\ , \ \
 \Delta_{sc}\simeq\Delta_{tr}^b\left(\frac{R_{pl}}{R_*}\right)\frac{1}{\sin\theta}\, .
\end{equation}
Combining these expressions we get
\begin{equation}
{\rm Small \  spot\!: \ \ \ \ \ }
TTV_{sc}
\simeq
-(1-\alpha) \mathcal{R}
\frac{\Delta_{tr}^b}{2}
\cos\phi_{sc}\sin\phi_{sc}
\ ,
\label{eq_TTV_impact_small}
\end{equation}

For the large spot case, the duration of the spot-crossing event,
$\Delta_{sc}$, is now
different, as the planet is crossing a spot which forms an
ellipse on the stellar disc, whose axes are $R_{spot}$ and
$R_{spot}\cos\psi$. One can show that the length of the planet's
path on the spotted area is
$R_{spot}\sqrt{\cos^2\theta+\sin^2\theta\cos^2\phi_{sc}}$. We
therefore get
%
\begin{equation}
{\rm Large \ spot\!: \ \ \ \ \ }
\delta_{sc}\simeq(1-\alpha)\left(\frac{R_{pl}}{R_*}\right)^2
\ \ , \ \
 \Delta_{sc}\simeq\Delta_{tr}^b\left(\frac{R_{spot}}{R_*}\right)
\sqrt{\cot^2\theta+\cos^2\phi_{sc}} \, ,
\end{equation}
%
and thus
\begin{equation}
{\rm Large \  spot\!: \ \ \ \ \ }
TTV_{sc}
\simeq
-(1-\alpha) \mathcal{R}
\frac{\Delta_{tr}^b}{2}
\sqrt{\cot^2\theta+\cos^2\phi_{sc}}\
\sin\phi_{sc}
\ ,
\label{eq_TTV_large}
\end{equation}

\newpage

We can see that for a non-vanishing impact parameter there is a
difference between the large and small planet cases, unlike in
the basic model.
The difference is due to the $\cot^2 \theta$ term under the
square sign in Equation~(\ref{eq_TTV_large}).
Note that the approximation of the large spot is not valid for
$|\phi|\simeq\pi/2$, where the projected area of the spot is
small. Hence, we inserted into the calculation of the large-spot
case a correction factor that turns the TTV expression to be
similar to the
small-spot one when $|\phi|\rightarrow\pi/2$.
This was done by multiplying the $\cot^2 \theta$ term with a
Fermi function that is approximately unity, except for
$|\phi|\rightarrow\pi/2$, when the correction factor goes to
zero.
%

\subsubsection{Limb darkening and impact parameter}
%
To further extend our simplistic model, we consider now a case
for non-zero impact parameter {\it and} quadratic limb darkening
together.
As before, we divide the discussion between the cases of small
and large spot.
Following Equation~(\ref{eq_TTV_impact_small}), but now
multiplying it by the limb darkening brightness factor, we get
for the small spot case:
%
\begin{equation}
\begin{split}
{\rm Small \  spot\!: \ \ \ \ \ }
TTV_{sc} & \simeq -(1-\alpha) \mathcal{R} \frac{\Delta_{tr}^b}{2}
\big\{ (1- g_1 -g_2) \sin\phi (t) +  \\
&
+ (g_1+2g_2) \sin \theta \sin \phi (t) \cos \phi (t) - \\
&
-g_2\sin ^2\theta \sin \phi (t) \cos^2 \phi (t) \big\}  \cos
\phi(t) \, ,
\end{split}
\end{equation}
while for the large spot case, following
Equation~(\ref{eq_TTV_large}), we get:
%
\begin{equation}
\begin{split}
{\rm Large \  spot\!: \ \ \ \ \ }
TTV_{sc}
&
\simeq -(1-\alpha) \mathcal{R} \frac{\Delta_{tr}^b}{2} \big\{ (1-
g_1 -g_2) \sin\phi (t) +
\\ &
+(g_1 +2g_2) \sin \theta \sin \phi (t) \cos \phi (t) -
\\&
- g_2\sin ^2\theta \sin \phi (t) \cos^2 \phi (t) \big\} \sqrt{
\cot ^2 \theta  + \cos^2 \phi(t)} \, .
\end{split}
\label{eq:impact_parameter}
\end{equation}

\subsubsection{Stellar obliquity}
\label{sec:obliquity}

The last case we consider is when the apparent planetary chord
along the stellar disc goes through the center ($b_{pl}=0$), but
is inclined with the angle $\eta$ relative to the
stellar equator.
We nevertheless assume that in some transits spot-crossing events
happen, with spots that have different latitudes. In such cases,
$t_{sc}$ is proportional to the  distance of the spot-crossing
event from the center of the disc, as in the basic model
(Equation (3)).
Similar considerations show that here also we get, as in Equation
(12):
%
%
%
\begin{equation}\nonumber
TTV_{sc}
\simeq
-(1-\alpha) \mathcal{R}
\frac{\Delta_{tr}}{2}
\cos\psi_{sc}\sin\psi_{sc}
\ ,
\label{eq_TTV}
\end{equation}
%
which is true for small and large spot cases alike. The extension
for limb darkening also holds in this case.

\subsection{Comparing the different TTV patterns}
\begin{figure}[p!] 
\centering
\resizebox{16cm}{12cm}
{\includegraphics{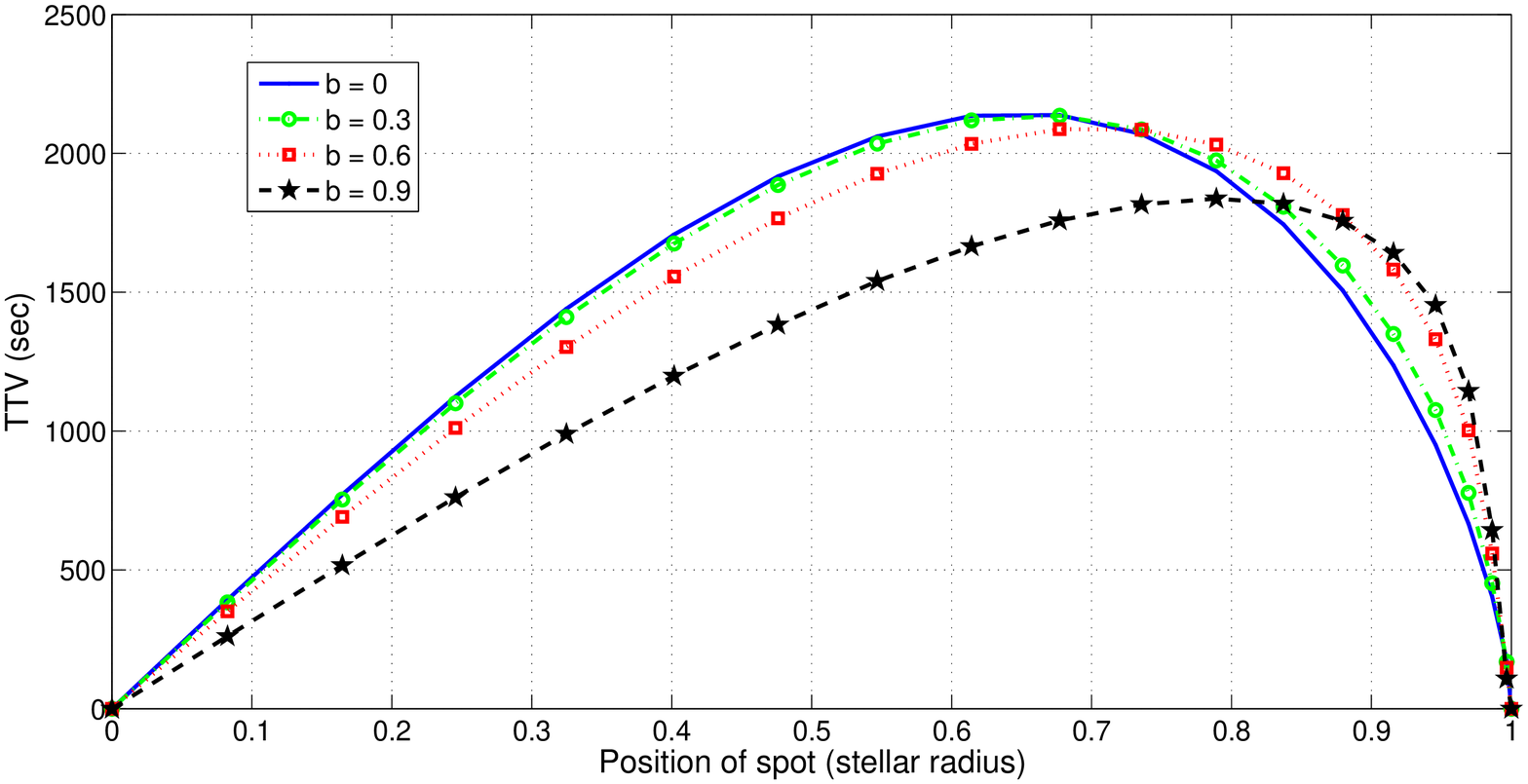}}
   \caption{The analytic approximation for the induced TTV as a
function of the spot-crossing position on the stellar disc for
different values of the impact parameter, using the large-spot
expression of Equation (\ref{eq:impact_parameter}).
The position of the spot-crossing event is measured relative to
the center of the stellar disc, in units of the stellar radius.
The graphs are for
a Jupiter-size planet that orbits a star with solar radius in a
3-d orbit. The duration of the transit (mid-ingress to
mid-egress) is about 2.62 hours, a value on which we based our
estimations. The spot and planet radii were chosen as
 $R_{spot}/R_*=0.15$ and $R_{pl}/R_*=0.05$.
The limb darkening coefficients used are [g$_1$, g$_2$] =
[0.29,0.34].
}
\label{fig:impact_parameter}
\end{figure}

To visualize the expected TTVs  derived by our  analytical
approximation for non-vanishing impact parameter cases, we
plotted in Figure \ref{fig:impact_parameter} the calculated TTVs
for different values of the impact parameter, with the large-spot
approximation, using $R_{spot}/R_*=0.15$ and $R_{pl}/R_*=0.05$
values.
We chose a typical parameters for a transiting system --- a
planet orbiting a star with solar radius in a 3 d orbit. The
duration of the transit (mid-ingress to mid-egress) is about 2.62
hours, a value on which we based our estimations.

One can see in the figure that the amplitude of the induced TTV
is about 5 min. The derived TTVs display almost linear slope as a
function of the spot-crossing position, up to a maximum at
distance of 0.6--0.85 stellar radii from the center of the
stellar disc, and then a sharp drop to zero at the edge of the
stellar disc.

%

\section{Comparison with numerical simulations}
\label{sec:simulations}

As noted in the introduction, \citet{boisse12} and
\citet{oshagh13a} developed a numerical tool ---
SOAP-T\footnote[1]{http://www.astro.up.pt/resources/soap-t/}, to
simulate stellar photometric modulations induced by a rotating
spot,  including a planetary transit light curve which includes a
spot-crossing event.
\citet{oshagh13b} used SOAP-T to derive detailed transit light
curves, and then fitted them with transit templates to obtain the
expected TTVs, very similar to what is performed when deriving
the TTVs from the \kepler\ actual data \citep[e.g.,][]{mazeh13}.
This is much more accurate derivation than that of the previous
section, where we estimated the TTVs by the center-of-light
approach.
It is therefore useful to compare the TTVs obtained by our
analytical approximation with the ones derived with the SOAP-T
numerical code and the transit fitting.

To do that we perform in this section two comparisons.
First, we used ourselves the publicly available SOAP-T tool to
produce transit light curves with spot-crossing events and fitted
them with the \citet{mazeh13} codes to produce TTVs for a few
cases and compare them with the analytical approximations.
Second, we derive  with our analytical center-of-light approach
some TTVs for the cases derived by \citet{oshagh13b}, and compare
the results.

\begin{figure}[t!]
 \includegraphics[width =
0.9\textwidth]{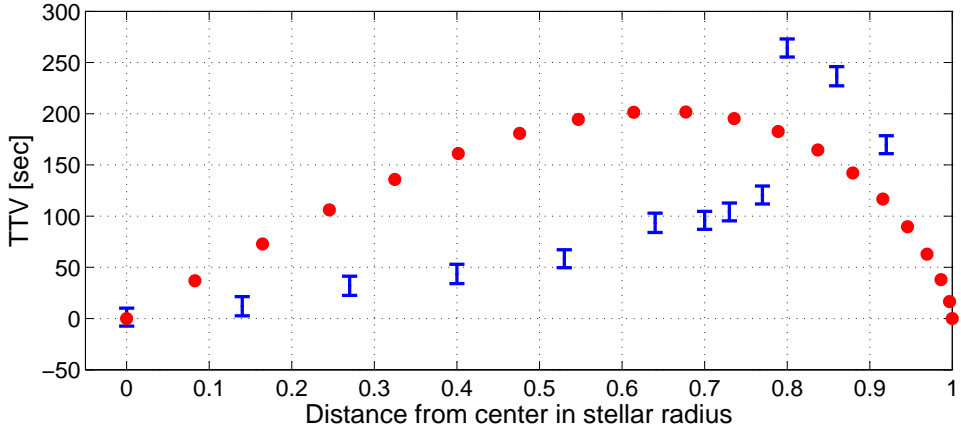}
 \caption{Comparison of the analytic approximation for the
induced TTV with numerical simulations, as a function of the
spot-crossing phase.
The approximated TTV (red) was derived by
Equation~(\ref{eq:TTV_basic_eq_LD}), while the light curves
obtained by the SOAP-T tool (blue) were analyzed to derive the
TTV. The error bars was derived from the \citet{mazeh13} codes.
$R_{pl}/R_*=0.1$ and $R_{spot}/R_*=0.1$. The limb darkening
coefficients that were used were [g$_1$, g$_2$] = [0.29,0.34].}
\label{fig:analytical_oshagh}
\end{figure}

In Figure~\ref{fig:analytical_oshagh} we plotted our analytical
approximation for the same system as before --- a 3 d transiting
planet orbiting a solar-like star. We used
limb darkening of $g_1=0.29$ and $g_2=0.34$, $R_{pl}/R_*=0.1$ and
$R_{spot}/R_*=0.1$,
a dark spot, with $\alpha=0$,
 and impact parameter of zero. We can see from the figure that
the maximum expected TTV based on our approximation is similar to
the one obtained when simulating the spot-crossing event. The
obvious difference is the phase dependence --- while the
analytical approximation has a smooth rise to the maximum, at
phase of 0.65, the simulated light curves yielded TTVs that are
quite small for most phases, and rise sharply towards the maximum
at phase 0.8.

%

The reason for this difference comes from the different
approaches of obtaining the TTV. The approach that fits a model
to the simulated light curve ignores sometimes the `bump' in the
light curve caused by the spot-crossing event, yielding a small
TTV, while the center-of-light model is, in fact, integrating
over the whole transit light curve. We will see this difference
again and again. Nevertheless, this difference does not change
the result of this paper --- the negative (positive) correlation
for prograde (retrograde) motion, as will be shown below.

\begin{figure}[t!]
\centering
\resizebox{15cm}{10cm}
{\includegraphics{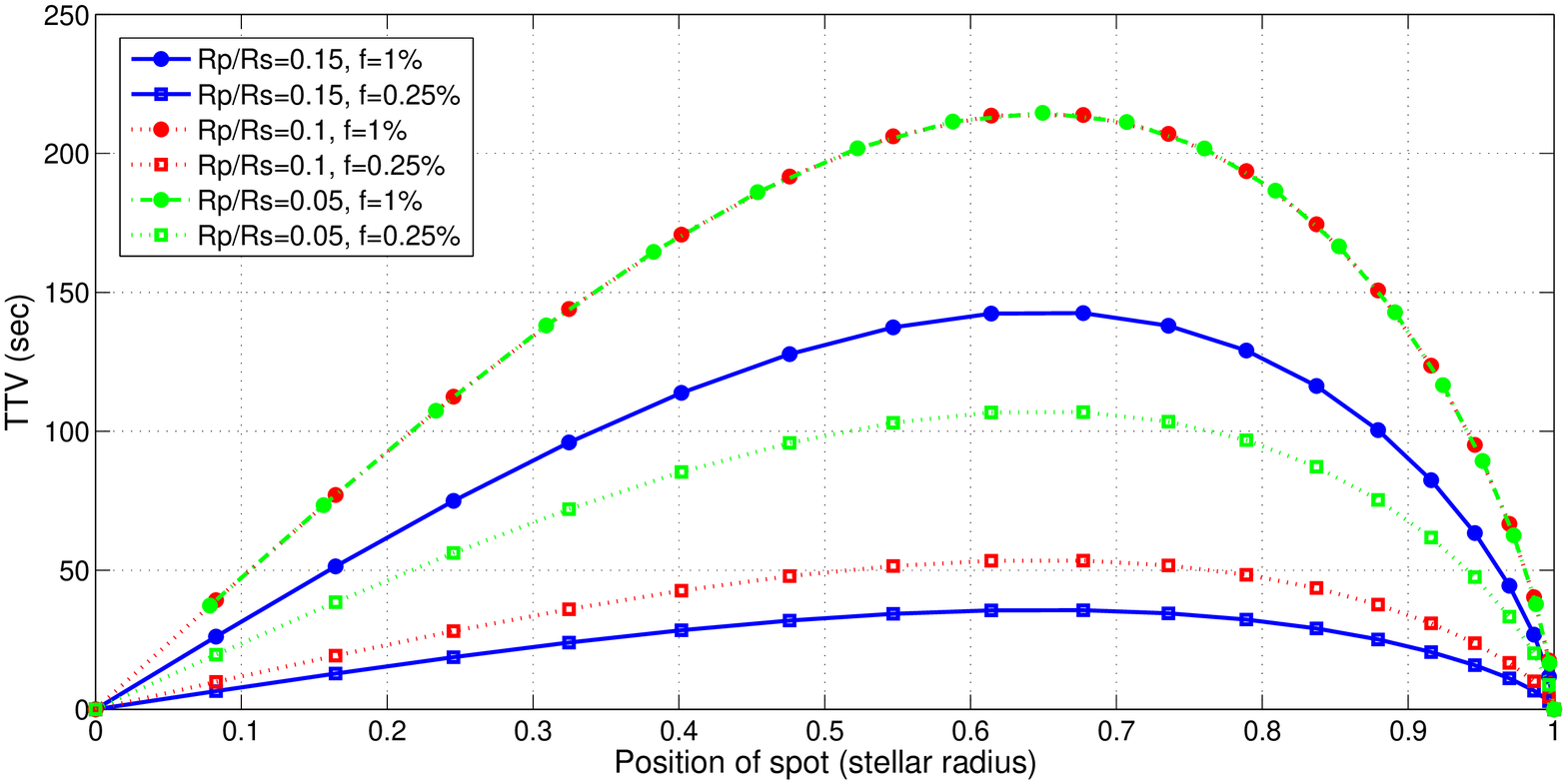}}
   \caption{The analytic approximation for the induced TTV as a
function of the spot-crossing phase for different spot and planet
sizes.
Expected TTV were derived by using
Equation~(\ref{eq:TTV_basic_eq_LD}).
Rp/Rs is planet to star radius ratio and f is spot to star radius
ratio squared. The limb darkening coefficients used are [g$_1$,
g$_2$] = [0.29,0.34].
}
\label{fig:oshagh1}
\end{figure}

\citet{oshagh13b} paper includes two figures that present their
derived TTVs as a function of the orbital phase of the
spot-crossing event. We applied our analytical approximation to
all cases included in  \citet{oshagh13b} figures, presented in
the next two figures.
In Figure \ref{fig:oshagh1} we plot results of our analytical
approximation that corresponding to the six cases of
 \citet{oshagh13b} Figure 3, where they have considered different
spot and planet relative sizes, keeping the same limb darkening
parameters.
We chose the same $(R_{spot}/R_*)^2$ (what they call 'f')
values --- 0.01 and 0.0025,
and the same  $R_{pl}/R_*$ values --- 0.05, 0.1, and 0.15.
We used the same limb darkening coefficients of [$g_1$,$g_2$] =
[0.29,0.34], and assumed a completely dark spot ($\alpha$ = 0 in
our notation).
As before, the transit duration is set to be 2.62 hours.

%

As in the previous figure, we see here that the maximum TTV is
similar to the values obtained by
\citet{oshagh13b}, while the phase behavior of the two approaches
is different, as explained above.

\begin{figure}[p!]
\centering
\resizebox{15cm}{10cm}
 {\includegraphics{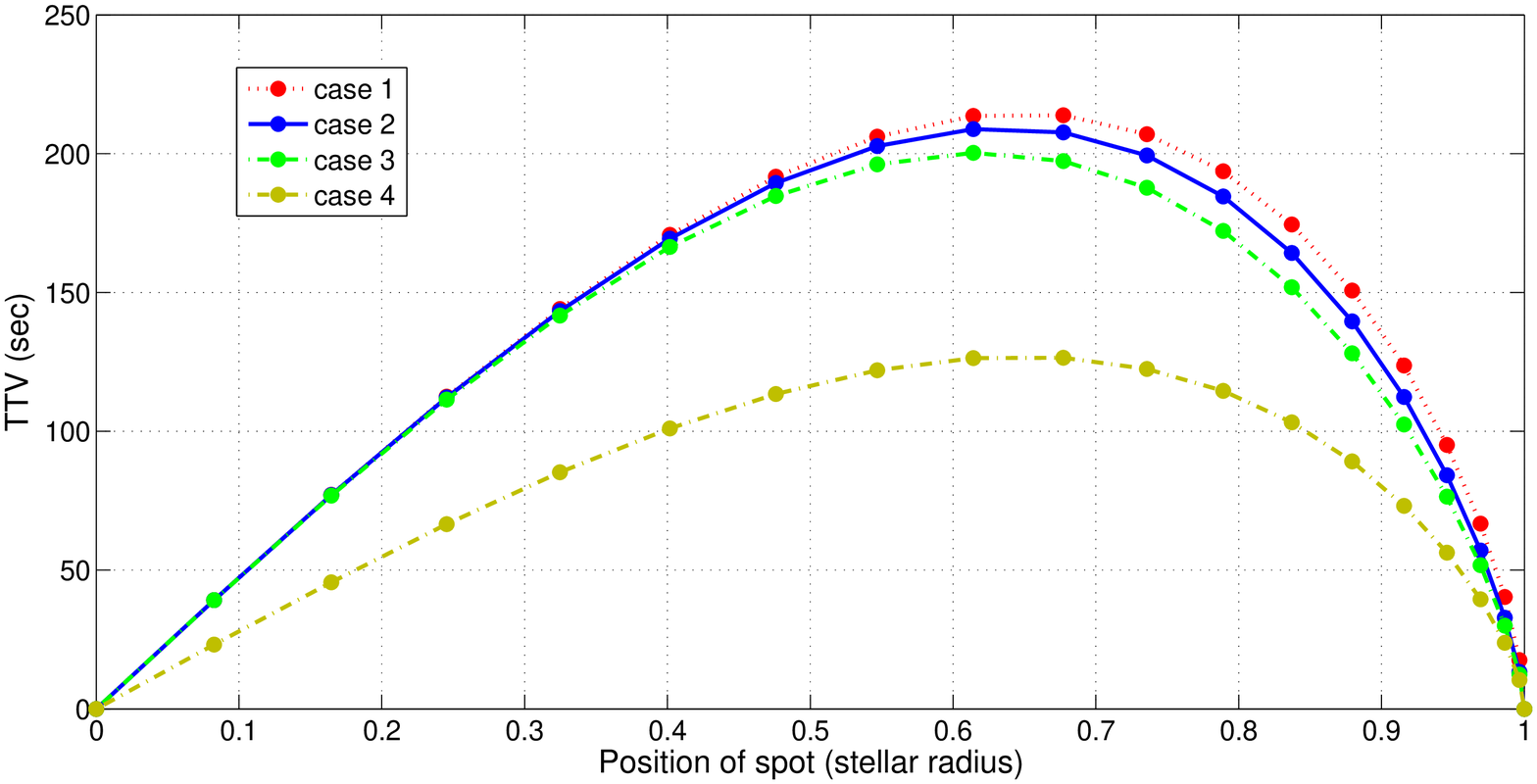}}
   \caption{Expected TTV for different limb darkening parameters,
using our analytical approximation for $R_{pl}/R_* =
R_{spot}/R_*=0.1$. The limb darkening coefficients were in
case 1 [g$_1$, g$_2$] = [0.29,0.34],  in case 2  [0.38,0.37], in
case 3 [0.6,0.16], and in case 4 [0.29,0.34]. In Case 4 the spot
has half of the stellar brightness ($\alpha = 0.5$), and the spot
size was increased by 1.4, in order to get similar amplitude of
the TTVs.}
\label{fig:oshagh2}
\end{figure}

Another comparison was done by constructing
Figure~\ref{fig:oshagh2} and comparing it with Figure 6 of
\citet{oshagh13b}, to study the effect of the limb darkening and
spot darkness. Here again the amplitudes of the analytical
approximation are similar to those of \citet{oshagh13b}, while
the phase dependence is different, like in our
Figure~\ref{fig:oshagh1}.

%

%
\section{Analytical approximation for the stellar photometric
slopes}
\label{sec:slopes}

We turn now to approximate the local photometric slope at the
time of the transit, assuming as before that the stellar
brightness is modulated by a single circular spot.

For no limb darkening and null impact parameter we
approximate the stellar flux, modulated by the spot as
%
\begin{equation}
F_*(t) \simeq 1-   \mathcal{A}\cos \psi(t)\, ,  \ \ \
{\rm for} -\pi/2\leq\psi\leq\pi/2 \, ,
\label{eq:stellar_photo}
\end{equation}
%
where $\mathcal{A}$ is the observed amplitude of the
photometric
modulation. This is so because the spot area on the stellar disc
is reduced by the aspect ratio $\cos\psi$.
The {\it derivative} of the stellar photometric brightness is
therefore
%
\begin{equation}
\label{fdot}
\dot{F}_*(t) \simeq \omega_{*} \mathcal{A}\sin \psi (t) \ .
\end{equation}

The amplitude of the observed stellar photometric modulation is a
function of the spot
radius and darkness. To express this relation we introduce the
$0<\beta<1$ parameter,
which accounts for the possibility that the spot crossed by the
planet might not be
the only spot that contributes to the stellar modulation with the
observed
phase. Therefore, $\beta$ measures the ratio of the area of the
spot being crossed by the
planet to the total neighboring spotted area that
causes the photometric modulation {\it with the same phase}.
The total stellar modulation due to the spots, relative to the
maximum
stellar brightness, is
%
\begin{equation}
 \mathcal{A}\simeq\frac{1-\alpha}{\beta}\left(\frac{R_{spot}}{R_*
}\right)^2 \, .
\end{equation}

In the case of limb darkening, the brightness of the spotted star
takes the form
%
\begin{equation}
F_*(t) \simeq 1-   \mathcal{A}\cos \psi(t)
\big\{1 -g_1(1-\cos\psi(t)) - g_2(1-\cos\psi(t))^2 \big\} \, ,
\label{eq:stellar_photo_LD}
\end{equation}
as the photometry is modulated by the aspect ratio and the limb
darkening at the spot's location.
The photometric derivative is then:
%
\begin{equation}
\dot{F}_*(t) =\mathcal{A}\omega_*
\big\{(1-g_1-g_2)\sin \psi (t) + (2g_1+4g_2) \sin \psi (t) \cos
\psi (t) -3g_2\sin \psi (t)\cos^2 \psi (t) \big\} \, .
\end{equation}

The stellar photometry for non-vanishing impact parameter is
expressed like in Equation~(\ref{eq:stellar_photo}), but now
$\cos \psi(t)=\sin \theta \cos \omega_*t$,
 and therefore the stellar photometry derivative is
%
%
\begin{equation}
\dot{F}_*(t) \simeq \omega_{*} \mathcal{A}\sin\theta\sin \phi (t)
\, ,
\label{eq:fdot_impact}
\end{equation}
where $t$ is the time since the {\it spot} was in the middle of
its trail, on the projection of the stellar spin (see below) on
the stellar disc, and $\omega_*$ is the stellar rotation rate, as
explained in Section~\ref{sec:obliquity}.

For non-vanishing impact parameter {\it and} stellar limb
darkening the stellar photometry is
\begin{equation}
F_*(t) \simeq 1-   \mathcal{A} \sin \theta\cos \phi(t)\
\big\{1 -g_1(1- \sin \theta\cos\phi(t)) - g_2(1-\sin
\theta\cos\phi(t))^2 \big\} \, .
\end{equation}
%
and its derivative is
%
\begin{equation}
\begin{split}
\dot{F}_*(t) \simeq
&
 \omega_{*} \mathcal{A}\big\{(1-g_1-g_2)\sin\theta\sin \phi (t) +
(2g_1 +4g_2)\sin ^2\theta \sin \phi (t) \cos \phi (t)
\\&
 - 3g_2\sin ^3\theta \sin \phi (t) \cos^2 \phi (t) \big\}\, .
\end{split}
\label{eq:fdot_impact_limb}
\end{equation}

When the obliquity of the system is non-vanishing, the spot moves
on a  chord orthogonal to the projection of the stellar
rotational axis, at a
colatitude $\theta_{spot}$, with $b_{spot}=cos\theta_{spot}$.
The spot chord is different from that of the planet, which we
assume goes through the center of the stellar disc.
Because of the inclination of the transit chord, at the time of
crossing
%
\begin{equation}
\sin\theta_{spot}\sin\omega_*t=\sin\psi_{sc}\cos\eta
\end{equation}
where $t$ is the time since the {\it spot} was in the middle of
its trail, on the projection of the stellar spin on the stellar
disc, and $\omega_*$ is the stellar rotation rate. Therefore, the
stellar photometric derivative is like
Equations~(\ref{eq:fdot_impact}) or (\ref{eq:fdot_impact_limb}),
except for a $\sin\theta_{spot}$ factor.
Note that when $\eta\rightarrow\pi/2$ then
$\dot{F}_*(t)\rightarrow0$, because the spot-crossing effect
occurs near the photometric maximum, and therefore the
correlation with the TTVs becomes difficult to detect.
%

\section{The correlation between $TTV_{sc}$ and the stellar
photometric slopes}
\label{sec:correlations}

We are now ready to consider the expected correlation between the
TTVs induced by the spot-crossing events and the local slope of
the stellar photometry at the time of the transit.

\subsection{TTV as a function of the photometric slope}

\begin{figure}
\centering
\resizebox{15cm}{10cm}
{\includegraphics{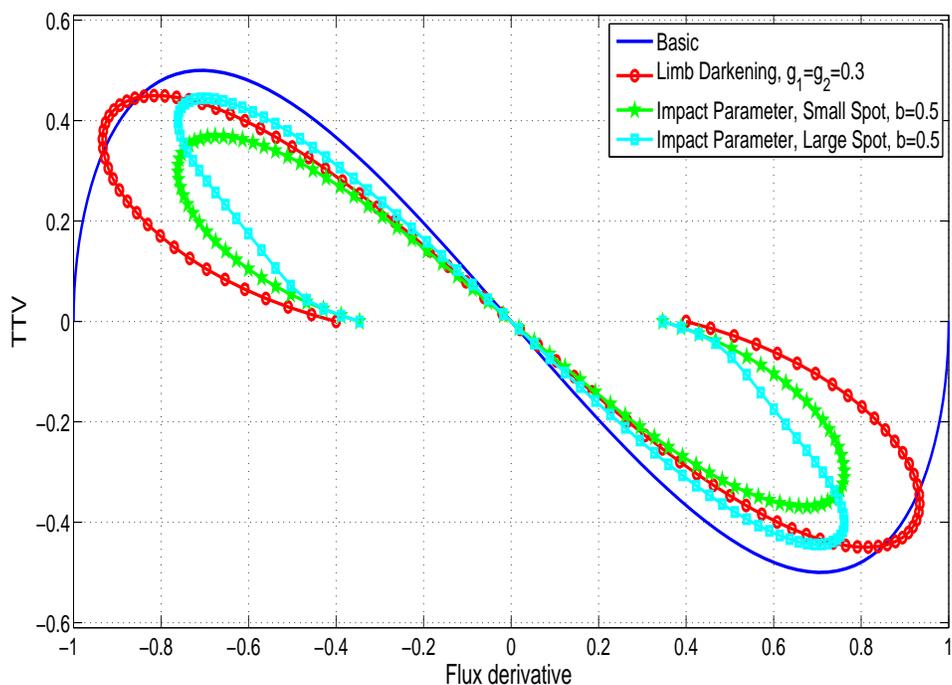}}
\caption{
The induced TTV$_{sc}$ versus the photometric slope for {\it
prograde} motion, using arbitrary units on both axes.
The blue line is the basic model, for $b=0$ and no limb
darkening. The red line presents the limb-darkening,
$g_1=g_2=0.3$, model, the green one is for $b=0.5$ and small
spot, and the cyan line is for the same $b$ with the large spot
approximation.
}
\label{fig:ttvfdot}
\end{figure}

Figure~\ref{fig:ttvfdot} displays our analytical approximation
for the TTVs as a function of the photometric derivatives for a
few cases.
The figure shows that the slope
of the stellar brightness at the time of each transit and the
corresponding induced TTV have {\it opposite} signs for prograde
motion, and therefore we expect negative correlation between the
two.
Obviously, the slope and the induced TTV have the {\it same}
sign for retrograde motion, because of the argumentation
presented in Section~2 and plotted in Figures~\ref{fig:prograde}
and
\ref{fig:retrograde} still holds, and therefore a positive
correlation is expected in such a case.

%

\subsection{Correlation as a function of noise and number of
observed transits}
\begin{figure}[pt!]
\centering
\resizebox{15cm}{10cm}
{\includegraphics{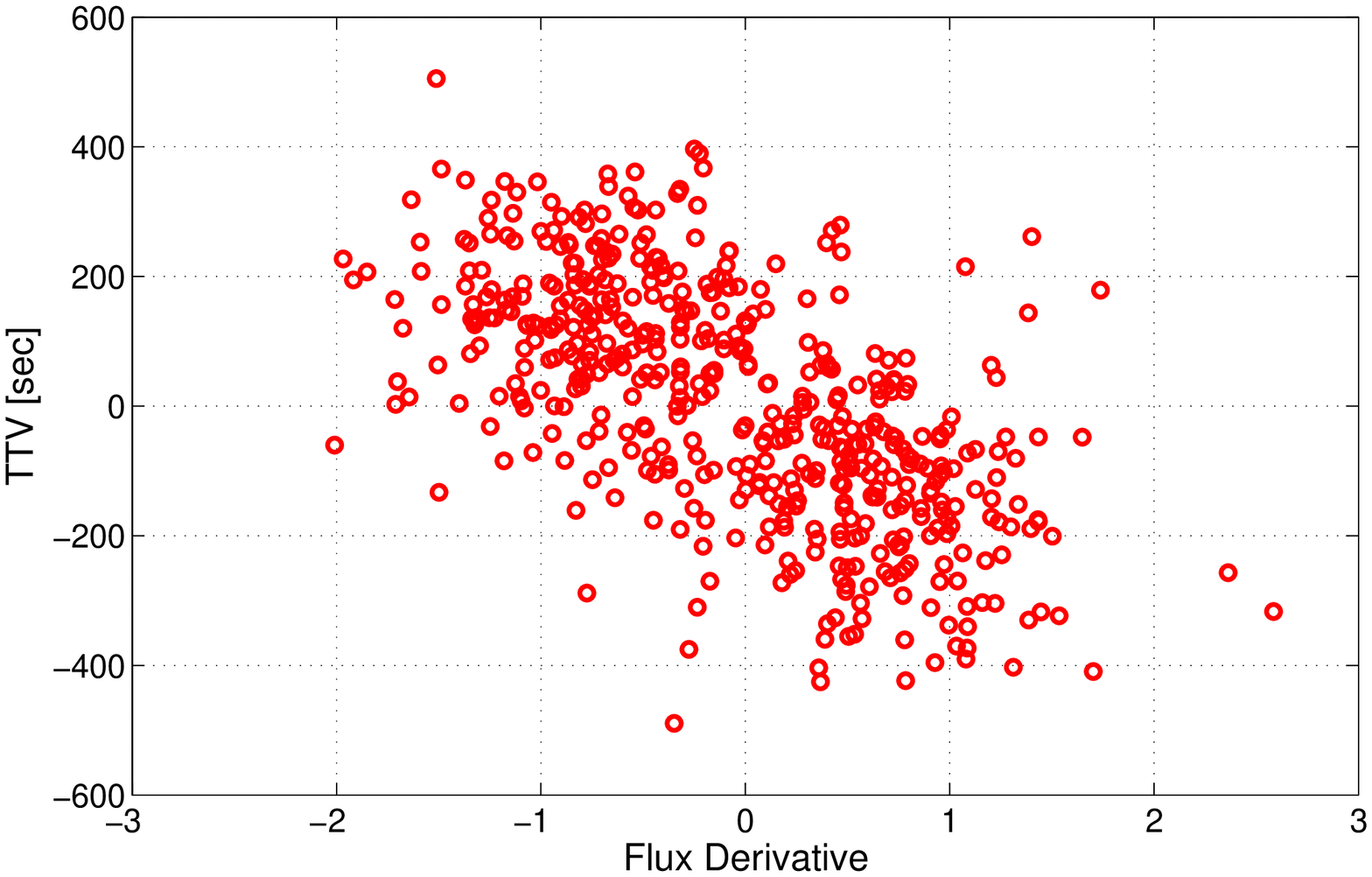}}
\caption{
Simulation of TTV, derived by the analytical approximation,
versus the corresponding photometric slope for {\it
prograde} motion, both with added normally distributed random
noise.
The noise r.m.s.~equals to 50\% of the maximum of the
corresponding variable.
The slope is scaled such that its maximum (before adding the
noise) is unity. The simulation includes 500 phases selected at
random. Correlation is $-0.62$. See text for details.
}
\label{fig:ttvvsdotanal}
\end{figure}

\begin{figure}[pt!]
\centering
\resizebox{15cm}{10cm}
{\includegraphics{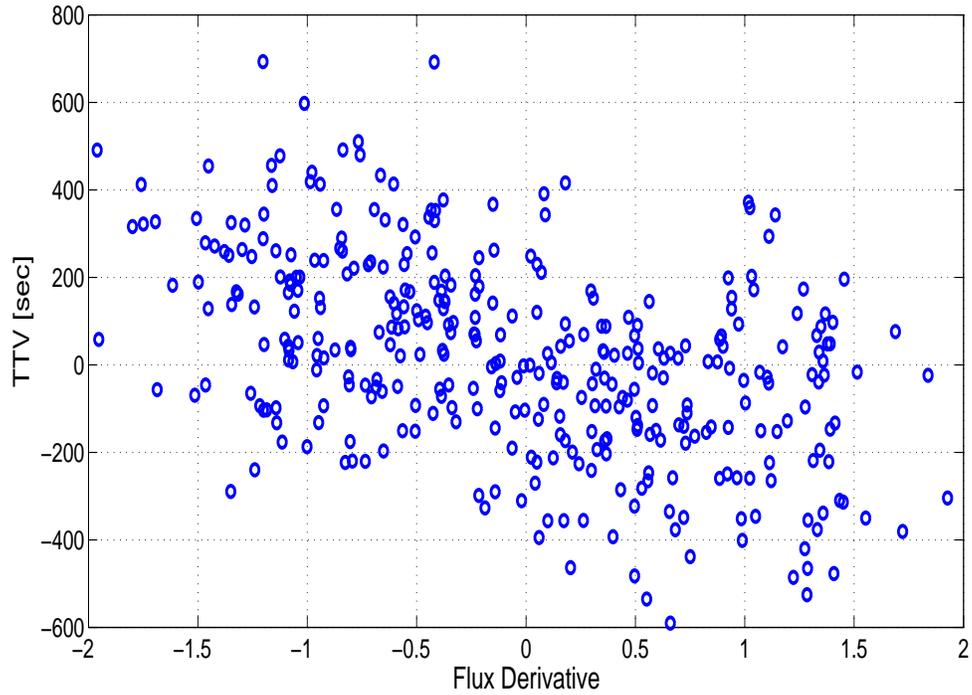}}
\caption{
Simulation of TTV, derived by analyzing the transit light curves
obtained by the SOAP-T tool, versus the corresponding photometric
slope for {\it prograde} motion, both with added normally
distributed random noise. Correlation is $-0.48$. See
Figure~\ref{fig:ttvvsdotanal} and text for details.
}
\label{fig:ttvvsdotoshagh}
\end{figure}

Figure~\ref{fig:ttvfdot} portraits how the TTVs derived by our
analytical approximation depend on the photometric slope, but it
does not show the real expected TTV, nor includes any
observational noise, associated with every derived TTV and
photometric derivative series. To see how these two affect the
expected correlation we added normally distributed noise to both
the simulated TTVs and the photometric derivatives, the results
of which are plotted in Figure~\ref{fig:ttvvsdotanal} for our
analytically approximated TTVs, and in
Figure~\ref{fig:ttvvsdotoshagh} for  TTVs derived by the SOAP-T
tool.

In both figures we used the same fiducial system, but now with
$R_{spot}/R_* = R_{pl}/R_* = 0.1$ and $[g_1,g_2] = [0.29,0.34]$.
The photometric derivative was scaled so that its maximum was
unity.
We chose at random 500 phases to be plotted in the figures, and
added randomly distributed normal noise
 to both the TTVs and the slope derivatives.
The noise r.m.s.~was equal to 50\% of the maximum of the
corresponding variable.
This amounts to 150 sec error on the TTVs and $0.5$ to the scaled
slope.
%

%
%

%
\newpage

The two figures show similar results --- there is a very clear
anti-correlation between the induced TTVs and the photometric
slopes at the transit timings, even when some small noise is
added. In fact, the noise covers up the fact that for some phases
the dependence of the TTVs on the slope changes its sign, as we
see in Figure~\ref{fig:ttvfdot}.

\begin{figure}[t!]
\centering
\resizebox{17cm}{10cm}
{\includegraphics{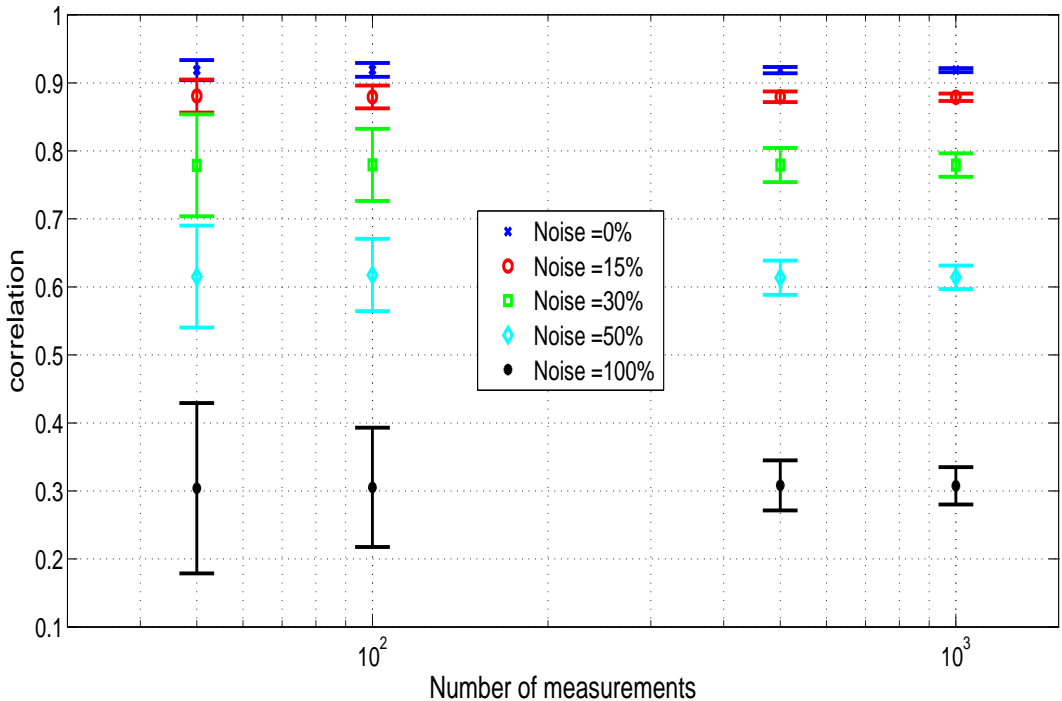}}
\caption{
The absolute value of the correlation of 1000 system samples of
simulated induced TTV with the stellar photometric slopes, for
different noise levels and different number of points.  The
points are the median of each sample and the error bars are the
sample r.m.s. See text for details.
}
\label{fig:correlation_noise}
\end{figure}

To estimate the expected effect of the noise on the measured
correlation we ran extensive simulations,
with different values of noise level and number of observed
transits.
For each choice of noise level, $\sigma_{\rm TTV}$,
$\sigma_{\rm slope}$ and number of transits, $N$, we chose $N$
random phases, derived their TTVs and photometric derivatives,
added randomly distributed noise to both the TTVs and the stellar
photometric slopes, and then derived the resulting
(anti-)correlation. We repeated this simulation for 1000 times,
with the same values of noise level and number of points. We then
derived the median and scatter of the sample of correlations
obtained, which are plotted in Figure~\ref{fig:correlation_noise}
as a function of the noise level and $N$.

We chose five values for $\sigma_{\rm TTV}$ and
$\sigma_{\rm slope}$, each scaled as a fraction of the maximum of
its corresponding variable. The five noise-to-signal ratios we
chose were [0, 0.15, 0.3, 0.5 1]. Each choice characterizes both
the noise added to the TTVs and to the photometric slopes. For
$N$ we chose values of 50, 100, 500, and 1000. For short-period
transiting planets \ik light curves could have on the order of
1000 transits, but 200--400 was a more typical number. All
together we had $4\times5=20$ sets of simulations, the results of
which are plotted in Figure~\ref{fig:correlation_noise}.

%

The expected value of the correlation depends on the noise level.
It goes from 0.9 for no noise down to 0.3 for a SNR of unity.
The $1\sigma$ spread of the correlation depends on the noise
level and the number of points. It goes from 0.13 for $N=50$ and
SNR of unity down to 0.02 for
$N=1000$ and no noise. The figure suggests that we can easily
detect the correlation with SNR of unity, if we can measure on
the order of 500 TTVs and their corresponding photometric slopes.


\section{Discussion}
\label{sec:discussion}

We presented here a simple approach that can, in a few cases, use
the
derived TTVs of a transiting planet to distinguish between a
prograde and
a retrograde planetary motion with respect to the stellar
rotation, assuming the TTVs are induced by spot-crossing events.
Using a simplistic analytical approximation we showed that those
TTVs might
have negative (positive) correlation with the local stellar
photometric slopes at the transit
timings for prograde (retrograde) motion. We have shown that the
correlation might be detected for different stellar limb
darkening and different impact parameters. Furthermore, we
obtained similar correlated TTVs when we used the SOAP-T tool to
simulate transit light curves and derive the corresponding TTVs.
We have shown also that even if we include certain amount of
noise, the correlation is still detectable.

Can such a correlation surface above the observational noise?
The expected amplitude of the TTV can be estimated by
Equations~(\ref{eq_factor2})  and (\ref{eq_timing_max}). For
example,
a system with $R_{spot}\simeq R_{pl}\simeq 0.1 R_*$,
$(1-\alpha)/4\simeq 0.25$ and transit duration of 3 h
should show an induced TTV on the order of 5 min. So, we can
expect to observe the (anti-)correlation between the TTVs and the
photometric slopes only for systems with high enough
signal-to-noise ratio that allows timing precision of the order
of 5 min or better.
Note that for a 3 d transiting
planet in the \kepler\ field we have at hand data for up to about
400 transits,
enabling us to detect a correlation even if the noise is
comparable with the signal.

Obviously, the approximation and simulation presented here are
quite
simplistic.
First, spotted stars probably have more than one spot. The spot
eclipsed by the planet might not be the one dominating
the stellar flux modulation, and hence the local photometric
slope at the time of transit might be very different from the
expressions we developed here.
Note, however, that in our simulation we allowed an error of the
photometric slope that
can be as large as the slope itself, and showed that even in such
a case the correlation still can be detected.
Second, spots have different stellar latitudes, so some transits
might not have induced TTVs at all, contaminating the expected
correlation. To deal with this problem one might consider the
correlation of only the highly significant TTVs, which could show
the signal better.
Third, the system
obliquity can be very different from $0^{\circ}$ or
$180^{\circ}$, although most of the planets around cool stars,
with a temperature below about 6000\,K,
apparently are aligned with the stellar rotation
\citep{albrecht12,mazeh15}.
We have shown that for systems with non-vanishing obliquity and
null impact parameter the shape of the dependence of the TTV on
the photometric slope is the same, although the obliquity might
decrease or even eliminate  the correlation, because many
transits might not include a spot-crossing event at all.
Note, however, that even for significant obliquity the
correlation might still exist, assuming there will be enough
induced TTVs, probably caused by spots with different latitudes.
Here again one might ignore the non-significant TTVs when
searching for a correlation.
Fourth, the  observed transiting
system might have additional planets that induce dynamical TTVs,
completely shadowing the TTVs caused by spot crossing events.

Despite all these obstacles, the correlation studied here might
be solid enough
to show up for a few KOIs.
Although our method cannot give an accurate spin-orbit angle, but
can instead only indicate the sign of the orientation of the
planetary motion, the method might be useful nevertheless, as it
uses \ik long-cadence data that is publicly available
for all transiting planets. In the next paper (Holczer et
al., in preparation) we report on a search for
correlation between the available TTVs and the corresponding
local photometric slopes at the transit timings for all {\it
Kepler} KOIs, and  indeed find five convincing cases with
significant correlations.

The approach described here can in principle be applied to any
eclipsing system, whether it is  a transiting planet or a stellar
binary. For a binary system, the induced
observed minus calculated (O-C) eclipse timings can be estimated
with the small-spot approximation, for which the planetary radius
is that of the secondary. We therefore expect the TTVs  to be on
the same order of magnitude as for transiting planets.
However, as eclipses in binaries are usually deeper and longer
than the planetary
transits, we expect the O-Cs in eclipsing binaries to be more
precise.

In fact, a negative correlation between the O-Cs and the local
photometric slopes was identified already for the stellar
eclipsing binary
in the Kepler-47 circumbinary planet system \citep{orosz12}.
The authors detected O-C on the order of 1 min in the timing of
the primary eclipse, and used the derived linear trend to correct
the eclipse timings. The detection of a negative
correlation for Kepler-47 is consistent with a more detailed
analysis of the spot-crossing events, also done by
\cite{orosz12}, which indicates a prograde motion.

The method presented here
can be applied in the future to a large sample of systems
monitored by
current and
future space missions, like K2 \citep{howell14},
TESS \citep{ricker14}, and PLATO \citep{rauer14}, helping
discovering, without additional observations, interesting systems
that are worth following, and possibly find what are the
conditions for alignment or misalignment of stellar rotations and
orbital
motions of planets and stellar binaries.


\acknowledgments
We are grateful to the referee for very helpful comments that
helped us substantially improve the paper. We are thankful to the
authors of the SOAP-T tool that made it publicly available.
The research leading to these results has received funding from
the European Research Council under the EU's Seventh Framework
Programme (FP7/(2007-2013)/ ERC Grant Agreement No.~291352).
T.M. also acknowledges support from
the Israel Science Foundation (grant No.\,1423/11) and the
Israeli Centers of Research Excellence (I-CORE, grant
No.\,1829/12).
T.M. is grateful to the Jesus Serra Foundation Guest Program and
to Hans Deeg and Rafaelo Rebolo, that enabled his visit to the
Instituto de Astrof\'isica de Canarias, where the last stage of this research was completed.
This work was performed in part
at the Jet Propulsion Laboratory, under contract with
the California Institute of Technology (Caltech) funded
by NASA through the Sagan Fellowship Program executed
by the NASA Exoplanet Science Institute.



\begin{thebibliography}{}

\bibitem[Albrecht et al.(2007)]{albrecht07} Albrecht, S.,
Reffert, S., Snellen, I., Quirrenbach, A., \& Mitchell, D.~S.\
2007, \aap, 474, 565

\bibitem[Albrecht et al.(2009)]{albrecht09} Albrecht, S.,
Reffert, S., Snellen, I.~A.~G., \& Winn, J.~N.\ 2009, \nat, 461,
373

\bibitem[Albrecht et al.(2011)]{albrecht11} Albrecht, S., Winn,
J.~N., Carter, J.~A., Snellen, I.~A.~G.,
\& de Mooij, E.~J.~W.\ 2011, \apj, 726, 68

\bibitem[Albrecht et al.(2012)]{albrecht12} Albrecht, S., Winn,
J.~N., Johnson, J.~A., et al.\ 2012, \apj, 757, 18

\bibitem[Albrecht et al.(2014)]{albrecht14} Albrecht, S., Winn,
J.~N., Torres, G., et al.\ 2014, \apj, 785, 83

\bibitem[Barnes(2009)]{barnes09} Barnes, J.~W.\ 2009, \apj, 705,
683

\bibitem[Barnes et al.(2011)]{barnes11} Barnes, J.~W., Linscott,
E., \& Shporer, A.\ 2011, \apjs, 197, 10

\bibitem[Batalha et al.(2013)]{batalha13} Batalha, N.~M., Rowe,
J.~F., Bryson, S.~T., et al.\ 2013, \apjs, 204, 24

\bibitem[Batygin(2012)]{batygin12} Batygin, K.\ 2012, \nat, 491,
418

\bibitem[Boisse et al.(2012)]{boisse12}
Boisse, I., Bonfils, X., \& Santos, N.~C.\ 2012, \aap, 545, AA109

\bibitem[Chaplin et al.(2013)]{chaplin13} Chaplin, W.~J.,
Sanchis-Ojeda, R., Campante, T.~L., et al.\ 2013, \apj, 766, 101

\bibitem[Dawson(2014)]{dawson14}
Dawson, R.~I.\ 2014, \apjl, 790, LL31

\bibitem[Deming et al.(2011)]{deming11} Deming, D., Sada, P.~V.,
Jackson, B., et al.\ 2011, \apj, 740, 33

\bibitem[D{\'e}sert et al.(2011)]{desert11} D{\'e}sert, J.-M.,
Charbonneau, D., Demory, B.-O., et al.\ 2011, \apjs, 197, 14

\bibitem[Fabrycky \& Tremaine(2007)]{fabrycky07} Fabrycky, D., \&
Tremaine, S.\ 2007, \apj, 669, 1298

\bibitem[Fabrycky et al.(2012)]{fabrycky12} Fabrycky, D.~C.,
Ford,
E.~B., Steffen, J.~H., et al.\ 2012, \apj, 750, 114

\bibitem[Gizon \& Solanki(2003)]{gizon03} Gizon, L., \& Solanki,
S.~K.\ 2003, \apj, 589, 1009

\bibitem[Groot(2012)]{groot12} Groot, P.~J.\ 2012, \apj, 745,
55

\bibitem[Hartman et al.(2011)]{hartman11} Hartman, J.~D., Bakos,
G.~{\'A}., Noyes, R.~W., et al.\ 2011, \aj, 141, 166

\bibitem[Hansen(2012)]{hansen12} Hansen, B.~M.~S.\ 2012, \apj,
757, 6

\bibitem[H{\'e}brard et al.(2011)]{hebrard11} H{\'e}brard, G.,
Ehrenreich, D., Bouchy, F., et al.\ 2011, \aap, 527, L11

\bibitem[Hirano et al.(2012)]{hirano12} Hirano, T.,
Sanchis-Ojeda, R., Takeda, Y., et al.\ 2012b, \apj, 756, 66

\bibitem[Hirano et al.(2014)]{hirano14} Hirano, T.,
Sanchis-Ojeda, R., Takeda, Y., et al.\ 2014, \apj, 783, 9

\bibitem[Holt(1893)]{holt1893} Holt, J. R.\ 1893, \aap, 12, 464

\bibitem[Howell et al.(2014)]{howell14}
Howell, S.~B., Sobeck,
C., Haas, M., et al.\ 2014, \pasp, 126, 398

\bibitem[Irwin et al.(2009)]{irwin09} Irwin, J., Aigrain, S.,
Bouvier, J., et al.\ 2009, \mnras, 392, 1456

\bibitem[Mazeh et al.(2013)]{mazeh13} Mazeh, T., Nachmani, G.,
Holczer, T., et al.\ 2013, \apjs, 208, 16


\bibitem[Mazeh et al.(2015)]{mazeh15}
Mazeh, T., Peretz, H., McQuillan, A. et al.\ 2015, \apj,
submitted

\bibitem[McLaughlin(1924)]{mclaughlin24}
McLaughlin, D.~B.\ 1924, \apj, 60, 22

\bibitem[McQuillan, Mazeh \& Aigrain(2014)]{mcquillan14}
McQuillan, A., Mazeh, T., \& Aigrain, S.\ 2014, \apjs, 211, 24

\bibitem[Moutou et al.(2011)]{moutou11} Moutou, C., D{\'{\i}}az,
R.~F., Udry, S., et al.\ 2011, \aap, 533, A113

\bibitem[Naoz et al.(2011)]{naoz11} Naoz, S., Farr, W.~M.,
Lithwick, Y., Rasio, F.~A., \& Teyssandier, J.\ 2011, \nat, 473,
187

\bibitem[Nutzman et al.(2011)]{nutzman11} Nutzman, P.~A.,
Fabrycky, D.~C., \& Fortney, J.~J.\ 2011, \apjl, 740, L10

\bibitem[Orosz et al.(2012)]{orosz12} Orosz, J.~A., Welsh,
W.~F., Carter, J.~A., et al.\ 2012, Science, 337, 1511

\bibitem[Oshagh et al.(2013a)]{oshagh13a}
Oshagh, M., Boisse, I., Bou{\'e}, G., et al.\ 2013a, \aap, 549,
AA35

\bibitem[Oshagh et al.(2013b)]{oshagh13b}
Oshagh, M., Santos, N.~C., Boisse, I., et al.\ 2013b, \aap, 556,
A19

\bibitem[Queloz et al.(2000)]{queloz00} Queloz, D., Eggenberger,
A., Mayor, M., et al.\ 2000, \aap, 359, L13


\bibitem[Rauer et al.(2014)]{rauer14}
Rauer, H., Catala, C., Aerts, C., et al.\ 2014, Experimental
Astronomy, 38, 249

\bibitem[Ricker et al.(2014)]{ricker14} Ricker, G.~R., Winn,
J.~N., Vanderspek, R., et al.\ 2014, \procspie, 9143, 91432

\bibitem[Rossiter(1924)]{rossiter24} Rossiter, R.~A.\ 1924, \apj,
60, 15

\bibitem[Sanchis-Ojeda et al.(2011)]{sanchis11a} Sanchis-Ojeda,
R., Winn, J.~N., Holman, M.~J., et al.\ 2011, \apj, 733, 127

\bibitem[Sanchis-Ojeda \& Winn(2011)]{sanchis11b} Sanchis-Ojeda,
R., \& Winn, J.~N.\ 2011, \apj, 743, 61

\bibitem[Sanchis-Ojeda et al.(2012)]{sanchis12} Sanchis-Ojeda,
R., Fabrycky, D.~C., Winn, J.~N., et al.\ 2012, \nat, 487, 449

\bibitem[Sanchis-Ojeda et al.(2013)]{sanchis13} Sanchis-Ojeda,
R., Winn, J.~N., Marcy, G.~W., et al.\ 2013, \apj, 775, 54

\bibitem[Schlaufman(2010)]{schlaufman10} Schlaufman, K.~C.\ 2010,
\apj, 719, 602

\bibitem[Schlesinger(1910)]{schlesinger1910} Schlesinger, F.\
1910, Publications of the Allegheny Observatory of the University
of
Pittsburgh, 1, 123

\bibitem[Shporer et al.(2012)]{shporer12} Shporer, A., Brown, T.,
Mazeh, T., \& Zucker, S.\ 2012, \na, 17, 309

\bibitem[Slawson et al.(2011)]{slawson11} Slawson, R.~W., Pr{\v
s}a, A., Welsh, W.~F., et al.\ 2011, \aj, 142, 160

\bibitem[Szabo et al.(2011)]{szabo11} Szabo, G.~M., et al.\ 2011,
\apjl, 736, 4

\bibitem[Szab{\'o} et al.(2013)]{szabo13} Szab{\'o}, R.,
Szab{\'o}, G.~M., D{\'a}lya, G., et al.\ 2013, \aap, 553, A17

\bibitem[Triaud et al.(2010)]{triaud10} Triaud, A.~H.~M.~J.,
Collier Cameron, A., Queloz, D., et al.\ 2010, \aap, 524, A25

\bibitem[Triaud et al.(2013)]{triaud13} Triaud, A.~H.~M.~J.,
Hebb, L., Anderson, D.~R., et al.\ 2013, \aap, 549, A18

\bibitem[Winn et al.(2006)]{winn06} Winn, J.~N., Johnson,
J.~A., Marcy, G.~W., et al.\ 2006, \apjl, 653, L69

\bibitem[Winn et al.(2010)]{winn10} Winn, J.~N., Fabrycky, D.,
Albrecht, S., \& Johnson, J.~A.\ 2010, \apjl, 718, L145

\bibitem[Winn et al.(2011)]{winn11} Winn, J.~N., Howard,
A.~W., Johnson, J.~A., et al.\ 2011, \aj, 141, 63

\end{thebibliography}
\end{document}